\documentclass[11pt]{article}

\usepackage[preprint]{acl}

\usepackage{times}
\usepackage{latexsym}

\usepackage[T1]{fontenc}

\usepackage[utf8]{inputenc}

\usepackage{microtype}

\usepackage{inconsolata}
\usepackage[hang,flushmargin]{footmisc}

\usepackage[breakable]{tcolorbox}
\usepackage{multirow}
\usepackage{graphicx}
\usepackage{adjustbox}
\usepackage{booktabs} 
\usepackage{amsmath}
\usepackage{xspace}
\usepackage{enumitem}

\title{Effective Dense Retrieval using Only In-Context Examples}

\author{
  \textbf{Nour Jedidi\textsuperscript{1}} \quad
  \textbf{Abdul Basit Ali\textsuperscript{1}} \quad
  \textbf{Hang Li\textsuperscript{2}} \quad
  \textbf{Jimmy Lin\textsuperscript{1}}
  \\[3pt]
  \textsuperscript{1}University of Waterloo \quad
  \textsuperscript{2}The University of Queensland
  \\[2pt]
  \texttt{njedidi@uwaterloo.ca}
}

\newcommand{\ours}{RICE\xspace} 

\begin{document}
\maketitle

\begin{abstract}
Turning decoder-only large language models (LLMs) into strong dense retrievers typically requires some form of retriever training. In this paper, we ask whether LLMs can instead be prompted to produce effective representations for dense retrieval given only a \emph{few} in-context examples. To answer this, we introduce \textbf{\ours} (\textbf{R}epresentations from \textbf{I}n-\textbf{C}ontext \textbf{E}xamples), a simple ``training-free'' approach that extracts high-quality dense representations from LLMs. To do so, \ours conditions the LLM on examples that provide a shared context for query and document encoding. Our results demonstrate that \ours embeddings can substantially improve the accuracy of prompt-based LLM embeddings, establishing it as a simple method to build LLM-based dense retrievers that do {\it not} require training. We release our code at \url{https://github.com/nourj98/RICE}.
\end{abstract}

\section{Introduction}

Large language models (LLMs) have emerged as powerful backbones for modern dense retrieval systems thanks to their rich world knowledge and text understanding capabilities~\cite{qwen3embedding}. But  ``out-of-the-box'', their hidden states do not automatically yield effective embedding representations for dense retrieval. As a result, strong LLM-based dense retrieval systems typically rely on additional retriever training, often through supervised~\cite{ma2024fine, thakur-etal-2025-hard}  or unsupervised~\cite{behnamghaderllm2vec,  behnamghader2026llm2vec} objectives. Building strong LLM-based embeddings that require \emph{no training} remains an open challenge. 

In this paper, we seek to address this challenge by drawing on two insights from the broader literature on zero- and few-shot dense retrieval. The first is that,  given only a few human-labeled in-context examples, LLMs can be prompted to produce effective \emph{task-specific} synthetic data for training downstream dense retrievers~\cite{daipromptagator, gwon2025study}. But as we aim to build LLM dense retrievers that do not require training, we draw on a second insight from PromptReps~\cite{zhuang2024promptreps}, which demonstrated that LLMs can be directly prompted to generate query and document representations for retrieval. However, one limitation of PromptReps is that query and document representations are derived independently via separate prompts, without any shared context connecting the two encoding tasks. 

Inspired by these insights, we ask: \emph{Can LLMs be prompted, using only a few in-context examples, to produce effective representations for dense retrieval?} To answer this, we introduce \textbf{\ours} (\textbf{R}epresentations from \textbf{I}n-\textbf{C}ontext \textbf{E}xamples), an approach for prompting LLMs to produce high-quality dense representations without requiring any form of retrieval-specific training. To do so, \ours combines the ease and simplicity of directly prompting LLMs for representations with the ability of LLMs to infer task-specific retrieval cues from only a few in-context examples.

Like PromptReps, \ours derives embeddings by prompting an LLM to generate a representative word for a query or document and extracting its final hidden state. Unlike PromptReps, which relies on an instruction alone, \ours primes this word generation process using query–document pairs that illustrate how the LLM \emph{itself} represents the given document through PromptReps-generated words. These demonstrations provide a shared context for query and document encoding with the aim of producing representations whose inner products better reflect relevance.

In summary, our contributions are twofold. First, we propose \ours, a simple approach for extracting -- via prompts -- dense retrieval representations from LLMs using \emph{only} in-context examples. Second, we comprehensively evaluate \ours on various retrieval datasets from BEIR~\cite{thakur2021beir}, and show that with only a \emph{few} in-context examples, LLMs can produce strong representations for dense retrieval without requiring retrieval-specific training. In particular, we demonstrate that \ours improves the effectiveness of prompt-based LLM representations. When compared against other dense retrieval approaches that also work directly  “out-of-the-box”, like HyDE~\cite{gao2023precise},  \ours achieves the highest average accuracy among these evaluated approaches. When compared against LLM2Vec-Gen~\cite{behnamghader2026llm2vec}, a highly effective self-supervised dense retriever, \ours is competitive. While \ours is less effective than dense retrievers trained on large collections of high-quality data, our findings show that in-context examples can offer a promising approach for narrowing the effectiveness gap between training-free, prompt-based retrieval and supervised dense retrievers.
\section{Methodology}

In this section, we first review the dense variant of PromptReps (PromptReps-Dense). We then introduce \ours, a method that prompts LLMs to produce dense representations conditioned on in-context exemplars.

\subsection{Preliminaries: PromptReps-Dense}
\label{sec:PromptReps}
The core idea behind PromptReps~\cite{zhuang2024promptreps} is that LLMs can be directly prompted to generate  effective representations for retrieval. For PromptReps, this is achieved by extracting the final hidden-state representation that the LLM produces when it is prompted to generate a representative word which summarizes a given input text. 

More formally, PromptReps goes as follows: Given some input text $x$ -- either a query $q$ or a document $d$ -- and a prompt
template -- which we denote as $P(\cdot)$ -- that asks the LLM for a
``single word that represents $x$ for a retrieval task'', PromptReps runs a
single forward pass over $P(x)$ and extracts the LLM's final hidden state immediately
before the language modeling head (i.e., the representation which produces the next-token distribution)
as the representation for retrieval:
\[
\begin{aligned}
\mathbf{e}_q &= \mathrm{LLM}_{\mathrm{hidden}}\bigl(P(q)\bigr). \\
\mathbf{e}_d &= \mathrm{LLM}_{\mathrm{hidden}}\bigl(P(d)\bigr).
\end{aligned}
\]

\noindent
To enable retrieval, $\mathbf{e}_d$ is pre-computed for each document in the corpus, $\mathcal{C}=\{d_1, d_2, \ldots, d_n\}$, and stored in an ANN index. At query time,  $\mathbf{e}_q$ is computed    ``on-the-fly'' and searched against the index.

We hypothesize that PromptReps-Dense has two potential limitations. 
The first is that, without task-specific relevance examples, the LLM lacks sufficient knowledge of what relevance ``looks like'' for a given search task, making it difficult to generate representative words that are well aligned with the retrieval objective. The second limitation is that PromptReps relies on an instruction alone, without any shared context to connect the query and document encoding tasks; each encoding task is performed independently without knowledge of how the other represents text, which may limit how well their inner products can capture relevance.

\subsection{\ours}
To overcome these challenges, we propose \ours. The core hypothesis underlying \ours is that just a \emph{few} in-context examples of query–document pairs, accompanied by PromptReps-generated representative words, can provide enough shared context to prime the LLM to produce query and document representations whose inner products better reflect relevance, while also exposing the model to task- and domain-specific language.

\ours operates as follows. First, building on the setup from \citet{daipromptagator}, we assume access to a few task-specific query-document pairs, $\{(q_1, d_1), (q_2, d_2),\ldots,(q_i, d_i)\}$, where $q_i$ denotes a \emph{human}-annotated query for document $d_{i}$. Next, for each document $d_i$ in the task-specific exemplar pairs, we use the PromptReps document encoding prompt to generate a representative word $w_i$. This yields a set of in-context exemplars, $f = \{(q_1, d_1, w_1), (q_2, d_2, w_2),\ldots,(q_i, d_i, w_i)\}$. These exemplars, $f$, are then incorporated into the \ours prompt template, $P_{\mathrm{RICE}}(f, x)$, as in-context  demonstrations, where each pair $(q_i, d_i)$ serves as an example user input, $w_i$ serves as the corresponding example assistant output, and $x$ is the query or document to be encoded:

\[
\begin{aligned}
\mathbf{e}_q^{\,f}
&= \mathrm{LLM}_{\mathrm{hidden}}\!\left(P_{\mathrm{\ours}}(f, q)\right), \\
\mathbf{e}_d^{\,f}
&= \mathrm{LLM}_{\mathrm{hidden}}\!\left(P_{\mathrm{\ours}}(f, d)\right).
\end{aligned}
\]

\noindent
As described in Section \ref{sec:PromptReps}, $\mathbf{e}_d^{\,f}$ is pre-computed  for  all documents in $\mathcal{C}$, and at query time $\mathbf{e}_q^{\,f}$ is generated and searched against the ANN index. 
\section{Experimental Setup}

Our experiments aim to study whether LLMs, using only in-context examples, can be prompted to produce effective representations for dense retrieval without the need for additional training.

\paragraph{Implementation} We implement \ours using Qwen3-8B~\cite{yang2025qwen3} and Qwen3.5-9B~\cite{qwen3.5}, with thinking disabled. 

To derive in-context examples, we use the Promptagator~\cite{daipromptagator, gwon2025study} setup, selecting examples from the training or development split when available and from the test split otherwise. For datasets in which the query–document pairs are drawn from the test set, when encoding a given query or document at inference time, we replace any in-context example that includes that query or document ``in-place'' with another sampled example that excludes it, so \ours is never exposed to a test judgment involving that query or document during its encoding. We leverage 10 in-context examples to generate $\mathbf{e}_q^{\,f}$ and $\mathbf{e}_d^{\,f}$. 

\paragraph{Datasets} We consider 10 retrieval datasets from BEIR~\cite{thakur2021beir}. The retrieval tasks include news retrieval (TREC-News, Robust04), financial question answering (FiQA),  biomedical IR (TREC-COVID, NFCorpus), fact checking (SciFact), citation prediction (SCIDOCS), tweet retrieval (Signal-1M),  argument retrieval (ArguAna), and question answering (NQ). For metrics, we report Recall@100 across all experiments.  

\paragraph{Baselines} Our primary point of comparison is the dense variant of PromptReps~\cite{zhuang2024promptreps} (PromptReps-Dense), which provides a baseline for measuring how much in-context examples improve LLM-prompted dense representations over a zero-shot approach. To ensure that our comparison between \ours and PromptReps accounts for differences in prompt wording, we include a \ours (Zero-Shot) baseline which uses the same query and document prompts as \ours, but without any in-context examples. Notably, \ours (Zero-Shot) differs from PromptReps only in the prompt text.

We next evaluate \ours against methods which utilize self-supervised or synthetic LLM-generated data to train a downstream dense retriever. The first method is LLM2Vec-Gen~\cite{behnamghader2026llm2vec}, a self-supervised approach for enabling decoder-only LLMs to produce dense representations by training them to represent a potential response to a given query rather than the query itself.  The second method is Promptodile~\cite{gwon2025study, daipromptagator}, which in-context prompts LLMs \emph{offline} to generate synthetic training data that is aligned with the target corpora for training dense retrievers. 

We then consider fully zero-shot dense retrieval methods where LLMs are leveraged as a tool to improve query representations for \emph{existing} unsupervised encoders. These include HyDE~\cite{gao2023precise}, which enriches the query representation using LLM-generated hypothetical answer documents; PRF-\textsc{Umbrela}~\cite{jedidi2026systematic}, which utilizes an LLM relevance judge to \emph{select} top-retrieved documents for refining the query representation; and CSQE~\cite{lei-etal-2024-corpus}, which refines the query using both LLM-generated hypothetical answer documents and an LLM relevance judge.\footnote{For CSQE, we use the \textsc{Umbrela}-HyDE adaptation from \citet{jedidi2026systematic}.} These baselines follow the implementation details described in \citet{jedidi2026systematic},  leveraging the unsupervised Contriever~\cite{izacardunsupervised} as the dense retriever with Rocchio vector feedback~\cite{li2022pseudo}. 

Lastly, we consider BGE-base-en-v1.5~\cite{xiao2024c} and Qwen3-Embedding-8B~\cite{qwen3embedding} as ``upper bounds'' on dense retriever effectiveness. Both systems have been fine-tuned using state-of-the-art dense retriever training pipelines on large amounts of data. BM25 is included as an unsupervised sparse retrieval baseline. 

\smallskip
\noindent
We utilize vLLM~\cite{woosuk2023vllm} for LLM inference, FAISS~\cite{douze2025faiss} for indexing and search, and Pyserini~\cite{lin2021pyserini} for its BM25 implementation and retrieval evaluation. 

\begin{table*}[t!]
\centering
\small
\setlength{\tabcolsep}{3pt}
\renewcommand{\arraystretch}{1.12}
\resizebox{\textwidth}{!}{%
\begin{tabular}{ll|cccccccccc|c}
\toprule
\textbf{Retriever}
& \textbf{LLM}
& \textbf{ArguAna}
& \textbf{FiQA}
& \textbf{News}
& \textbf{NFCorpus}
& \textbf{NQ}
& \textbf{Robust04}
& \textbf{SCIDOCS}
& \textbf{SciFact}
& \textbf{Signal-1M}
& \textbf{COVID}
& \textbf{Avg.} \\
\midrule

BM25 & --
& .932 & .539 & .447 & .246 & .751
& .375 & .348 & .925 & .370 & .109 & .504 \\

\midrule
\multicolumn{13}{l}{\textit{Dense Retrieval w/ Training}} \\
\midrule

BGE-base-en-v1.5 & --
& .992 & .742 & .499 & .337 & .942
& .351 & .496 & .967 & .311 & .141 & .578 \\

Qwen3-Embedding-8B & Qwen3-8B
& .996 & .929 & .570 & .388 & .977
& .494 & .636 & .973 & .292 & .194 & .645 \\

LLM2Vec-Gen & Qwen3-8B
& .988 & .670 & .484 & .310 & .871
& .356 & .430 & .948 & .245 & .122 & .542 \\

Promptodile & Llama-3.1-8B
& .994 & .709 & -- & .316 & --
& -- & .409 & .964 & -- & -- & -- \\

\midrule
\multicolumn{13}{l}{\textit{Dense Retrieval w/o Training}} \\
\midrule

HyDE
& \multirow{5}{*}{Qwen3-8B}
& .961 & .654 & .494 & .298 & .856
& .304 & .370 & \textbf{.964} & .219 & .090 & .521 \\
PRF-\textsc{Umbrela} &
& .909 & .568 & .439 & .304 & .765
& .282 & .346 & .909 & \textbf{.246} & .050 & .482 \\
CSQE &
& .952 & \textbf{.656} & \textbf{.516} & .312 & \textbf{.860}
& .337 & .371 & .952 & \textbf{.246} & .090 & .529 \\
PromptReps-Dense &
& .945 & .611 & .435 & .262 & .769
& .333 & .434 & .877 & .233 & .130 & .503 \\
\ours &
& \textbf{.967} & .655 & .503 & \textbf{.342} & .751
& \textbf{.361} & \textbf{.488} & .952 & .232 & \textbf{.135}
& \textbf{.539} \\

\midrule

HyDE
& \multirow{6}{*}{Qwen3.5-9B}
& .878 & .612 & .441 & .290 & \textbf{.848}
& .279 & .270 & .954 & .220 & .091 & .488 \\
PRF-\textsc{Umbrela} &
& .917 & .566 & .416 & .305 & .775
& .285 & .360 & .930 & .252 & .049 & .485 \\
CSQE &
& .920 & .640 & .467 & .305 & .845
& .333 & .323 & \textbf{.970} & .247 & .082 & .513 \\
PromptReps-Dense &
& .979 & .670 & .456 & .303 & .818
& .364 & .425 & .903 & \textbf{.279} & .140 & .534 \\
\ours (Zero-Shot) &
& \textbf{.980} & .689 & .435 & .305 & .688
& .354 & .457 & .962 & .219 & .128 & .522 \\
\ours &
& .975 & \textbf{.700} & \textbf{.512} & \textbf{.343} & .803
& \textbf{.407} & \textbf{.478} & .955 & .264 & \textbf{.145}
& \textbf{.558} \\

\bottomrule
\end{tabular}%
}
\caption{Main results (Recall@100) across BEIR datasets. Dense Retrieval w/ Training covers human labels, synthetic labels, and self-supervised objectives. Bold denotes the best result within each model family under Dense Retrieval w/o Training.}
\label{tab:main-recall100}
\end{table*}
\section{Results} 

Experimental results are shown in Table \ref{tab:main-recall100}. Overall, we find that \ours demonstrates strong effectiveness across BEIR tasks, achieving the highest average Recall@100 among the evaluated methods that require no training. 

In particular, \ours improves upon PromptReps-Dense with both Qwen3-8B and Qwen3.5-9B, yielding gains of 3.6 and 2.4 points in Recall@100, respectively, demonstrating that in-context examples improve the effectiveness of dense representations derived from prompting LLMs. The 3.6-point improvement of \ours (Qwen3.5-9B) over \ours Zero-Shot (Qwen3.5-9B) confirms that this improvement can be attributed to the inclusion of in-context examples rather than to differences between the \ours and PromptReps prompts.

Beyond methods which derive dense representations directly from LLMs, \ours is also more effective than HyDE, PRF-\textsc{Umbrela}, and CSQE, three approaches that instead utilize LLMs as a tool to improve query representations for \emph{existing} unsupervised encoders. Interestingly, this boost is larger when \ours utilizes Qwen3.5-9B instead of Qwen3-8B. One possible explanation, as argued by \citet{shen-etal-2024-retrieval}, is that such methods are constrained by the representational capacity of the unsupervised encoder. \ours does not have this bottleneck as it does not rely on a separate encoder and thus can directly take advantage of  improved LLM backbones.

When compared against methods trained using self-supervision or synthetic data, \ours achieves competitive retrieval effectiveness. In particular, when compared with LLM2Vec-Gen using the same Qwen3-8B backbone, \ours performs on par, on average. Using Qwen3.5-9B, \ours can improve upon LLM2Vec-Gen by 1.6 points in average Recall@100. Notably, for \ours, this improvement comes ``for free'', as it only requires another round of indexing, while LLM2Vec-Gen would need to undergo another round of self-supervised training. Across the five overlapping datasets, \ours achieves higher average Recall@100 than Promptodile, with gains on NFCorpus and SCIDOCS.

When evaluated against state-of-the-art dense retrievers that were trained using vast amounts of high-quality data (BGE-base-en-v1.5 and Qwen3-Embedding-8B), \ours lags behind.
Nevertheless, we highlight that our results indicate that \ours can provide a strong middle ground between approaches requiring no retriever training and those that rely on dedicated retriever training.
\begin{figure}[t]
  \centering
\includegraphics[width=0.9\columnwidth]{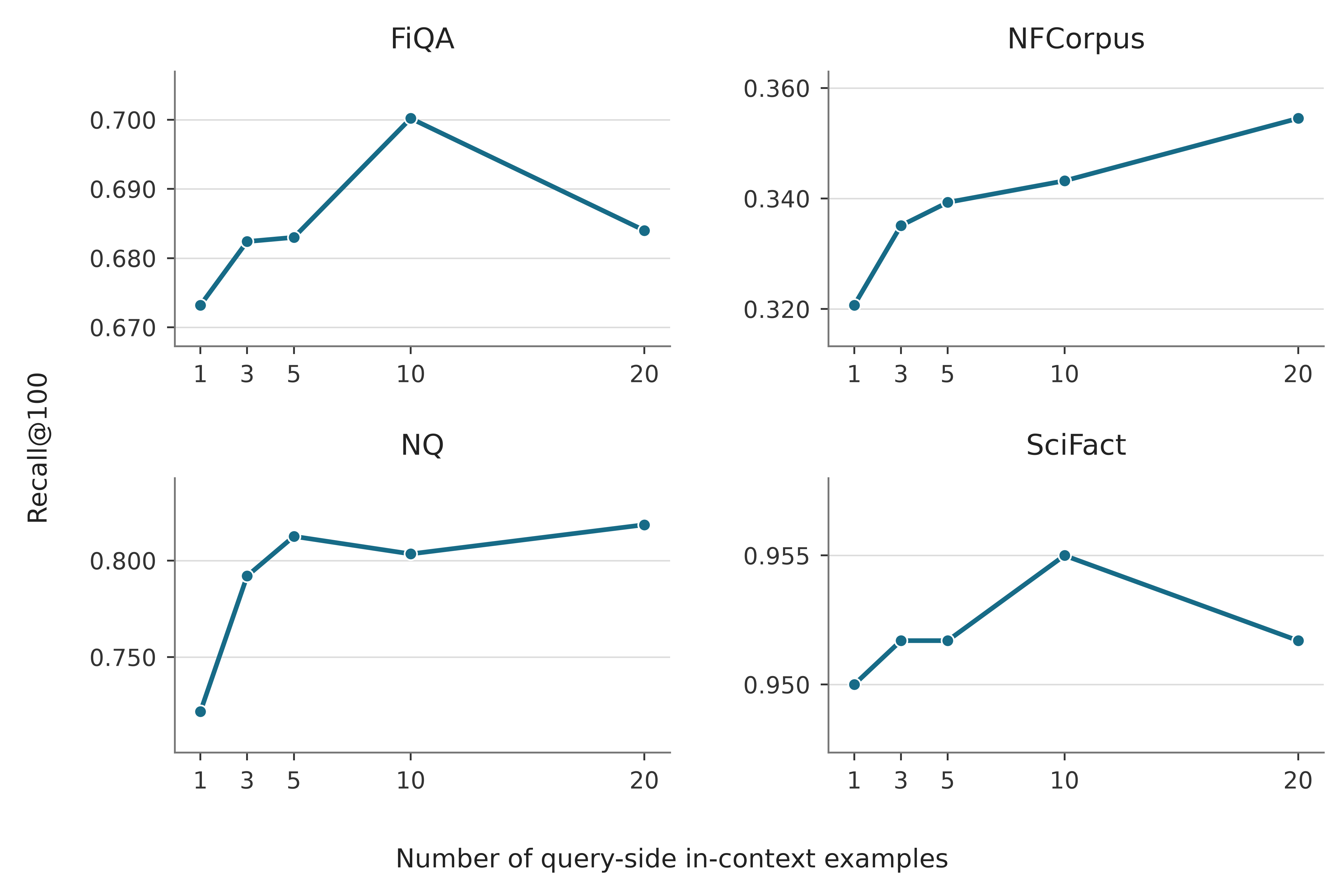}
  \caption{Recall@100 across varying numbers of in-context examples for generating \ours (Qwen3.5-9B) query encodings.}
  \label{fig:ndcg-varying-ic}
\end{figure}

\section{Analysis}

\paragraph{How many in-context examples are needed for query encoding?} 

In our experiments, we leverage 10 in-context examples for query and document encoding. We now investigate the effectiveness of \ours when varying the number of in-context examples used for generating $\mathbf{e}_q^{\,f}$. The results are in Figure~\ref{fig:ndcg-varying-ic}. Generally, the accuracy of \ours improves as the number of in-context examples increases, but can begin to decrease when utilizing 20 examples.

\paragraph{Dynamic exemplar selection for \ours.} In Table \ref{tab:query-specific-exemplars} we investigate whether dynamically selecting query- or document-specific examples can improve \ours. To study this, before encoding a query or document, we utilize BM25 to retrieve the top-10 most similar examples -- i.e.,  most similar queries for query-specific selection and documents for document-specific selection -- from the training exemplar pool, which then get paired with their representative word $w_i$ to construct $f$. 

The results show that dynamic in-context example selection can improve \ours's effectiveness across FiQA, NFCorpus, and SciFact. We note that 
the primary gains appear to come when utilizing dynamic exemplars for query encoding, and utilizing dynamic exemplars on only the document side appears to hurt effectiveness, compared to a fixed set of examples. 

\begin{table}[t!]
\centering
\small
\scalebox{0.9}{%
\begin{tabular}{l|ccc}
\toprule
\bf Dynamic Examples & \bf FiQA & \bf NFCorpus & \bf SciFact \\
\midrule
Fixed & .700 & .343 & .955 \\
\midrule
Query-Specific & .710 & .360 & .977 \\
Document-Specific & .672 & .333 & .933 \\
Query \& Document-Specific & .719 & .348 & .973 \\
\bottomrule
\end{tabular}%
}
\caption{Fixed and dynamic exemplar selection with \ours using Qwen3.5-9B.}
\label{tab:query-specific-exemplars}
\end{table}

\begin{table}[t!]
\centering
\small
\scalebox{0.9}{%
\begin{tabular}{l|cc}
\toprule
\textbf{Example Pairs} & \textbf{NFCorpus} & \textbf{SCIDOCS} \\
\midrule
None & .305 & .457 \\
\midrule
Relevant & .343 & .478 \\
Non-Relevant & .345 & .473 \\
Mixed & .345 & .484 \\
Random & .346 & .497 \\
\midrule
Relevant (cross-corpus) & .312 & .457 \\
\bottomrule
\end{tabular}%
}
\caption{\ours (Qwen3.5-9B) using different in-context example
compositions. None denotes \ours (Zero-Shot). Relevant
(cross-corpus) uses examples from Robust04.}
\label{tab:exemplar-composition}
\end{table}

\paragraph{Ablation on query-document exemplars.} Next, we investigate the composition of different in-context exemplars for \ours, exploring 
three settings: (1) \emph{Non-Relevant}, in which  documents judged relevant to one query are paired with different queries for which they are unjudged; (2) \emph{Mixed}, which combines five relevant and five non-relevant examples; and (3) \emph{Random}, where queries and documents are independently sampled from the query set and full corpus, respectively, while excluding judged query-document pairs. 

The results are in Table \ref{tab:exemplar-composition}. Interestingly, we find that all same-corpus variations of example pairs are stronger than using no in-context examples. Most surprising is that we see no clear advantage to using relevant example pairs versus the other alternatives.  

We believe an explanation for this result is that \ours  is effective because of the task- and domain-specific language present in the examples, and not necessarily the \emph{accuracy} of the examples. This would be consistent with the  findings from \citet{min-etal-2022-rethinking}, who showed that ground-truth demonstrations are not needed for effective in-context learning and that providing the LLM with in-distribution inputs is more critical.

To test this hypothesis, we explore using relevant cross-corpus demonstrations, i.e., relevant query-document pairs from another corpus, denoted by Relevant (cross-corpus) in Table \ref{tab:exemplar-composition}. We find this to be less effective than all variations of same-corpus example pairs, which further supports our hypothesis that the benefit of in-context demonstrations depends more on corpus-specific information than on query–document relevance alone: in-domain negative (non-relevant and random) demonstrations outperform relevant cross-corpus demonstrations.

\paragraph{Ablation on representative words.} The previous experiment suggests that selecting query-document pairs from the same corpus can matter more than their relevance relationships. Here, we further investigate this by studying the tradeoff between domain-specific language and the accuracy of in-context examples, probing the choice -- and importance -- of $w_i$. We investigate four variants: (1) \emph{Shuffled}, which permutes the original word labels across exemplars; (2) \emph{Fixed}, which assigns the same word to every exemplar; (3) \emph{Blank}, which replaces each word label with an empty string while retaining all demonstrations; and (4) \emph{Random}, which assigns ten distinct, unrelated words to the ten exemplars.

\begin{table}[t!]
\centering
\small
\scalebox{0.9}{%
\begin{tabular}{l|cc}
\toprule
\bf Representative Word & \bf NFCorpus & \bf SciFact \\
\midrule
Original & .343 & .955 \\
\midrule
Shuffled & .355 & .950 \\
Blank & .343 & .957 \\
Fixed: ``word'' & .337 & .958 \\
Fixed: ``orange'' & .326 & .928 \\
Fixed: ``feather'' & .345 & .948 \\
Random & .268 & .697 \\
\bottomrule
\end{tabular}
}
\caption{Representative word ablations for \ours with Qwen3.5-9B. Random  sets $w_1, \dots, w_{10}$ to feather, basket, curtain, candle, window, compass, marble, lantern, violin, and saddle.}
\label{tab:rep_word_shuffle}
\end{table}

The results can be found in Table \ref{tab:rep_word_shuffle}. We find that shuffling the word labels across the existing in-context examples or setting the labels to blank has little impact on the effectiveness of \ours. On the other hand, assigning distinct, unrelated words substantially reduces the effectiveness of \ours. Interestingly, using a fixed, random word has a less negative effect, although we note that its effect depends on both the choice of word and the dataset. One possible explanation for this could be that, when using a fixed word, there is no varying exemplar-specific signal, and the model can disregard the fixed word as ``noise''.

Together with the $(q, d)$ composition results, these findings suggest that the task- and domain-specific language in \ours is more critical than the actual accuracy of the in-context examples. In the case of selecting $w_i$, accurate exemplar-word associations are not the most essential component.
\section{Conclusion}

We introduce \ours, a simple approach for deriving effective  dense representations from LLMs using \emph{only} in-context examples. Empirical results demonstrate that \ours can improve the accuracy of prompt-based LLM dense retrievers while also outperforming other strong ``training-free'' dense retrieval baselines.  While fully supervised dense retrievers trained on large collections of high-quality relevance data remain more effective, our findings show that in-context examples can help narrow this gap.  We believe \ours represents an exciting direction for building dense retrieval systems that do not require retriever  training and can easily adapt to new tasks and domains. 

\section*{Acknowledgments}

This research was supported in part by the Natural Sciences and Engineering Research Council (NSERC) of Canada. Additional funding was provided by the Institute of Information \& Communications Technology Planning \& Evaluation (IITP) grant funded by the Korean Government (MSIT) (No.\ RS-2024-00457882, National AI Research Lab Project).


\bibliography{custom}

\clearpage
\appendix
\label{sec:appendix}
\newcommand{\promptTag}[1]{\textless#1\textgreater}
\newcommand{\promptImStart}{\promptTag{\textbar im\_start\textbar}}
\newcommand{\promptImEnd}{\promptTag{\textbar im\_end\textbar}}
\newcommand{\promptPlaceholder}{\{\}}
\newtcolorbox{promptbox}[1]{%
  colback=black!2,
  colframe=black!50,
  colbacktitle=black!8,
  coltitle=black,
  title={#1},
  fonttitle=\bfseries\footnotesize,
  fontupper=\ttfamily\footnotesize,
  boxrule=0.4pt,
  arc=1mm,
  left=1mm,
  right=1mm,
  top=1mm,
  bottom=1mm,
  before skip=0.75em,
  after skip=1em,
  before upper={\raggedright\setlength{\parindent}{0pt}\setlength{\parskip}{0pt}}
}









\begin{figure*}[t]
\centering

\begin{adjustbox}{
  max width=0.96\textwidth,
  max totalheight=0.78\textheight,
  keepaspectratio
}
\begin{minipage}{\textwidth}
\scriptsize

\begin{promptbox}{PromptReps query-encoding prompt}
\promptImStart{}system\\
You are an AI assistant that can understand human language.\\
\promptImEnd{}\\[0.15em]
\promptImStart{}user\\
Query: ``\promptPlaceholder{}''. Use one word to represent the query
in a retrieval task. Make sure your word is in lowercase.\\
\promptImEnd{}\\[0.15em]
\promptImStart{}assistant\\
The word is ``
\end{promptbox}

\begin{promptbox}{PromptReps document-encoding prompt}
\promptImStart{}system\\
You are an AI assistant that can understand human language.\\
\promptImEnd{}\\[0.15em]
\promptImStart{}user\\
Passage: ``\promptPlaceholder{}''. Use one word to represent the
passage in a retrieval task. Make sure your word is in lowercase.\\
\promptImEnd{}\\[0.15em]
\promptImStart{}assistant\\
The word is ``
\end{promptbox}

\begin{promptbox}{\ours{} query-side in-context example}
\promptImStart{}system\\
You are an AI assistant that can understand human language.\\
\promptImEnd{}\\[0.15em]
\promptImStart{}user\\
Query: ``\promptPlaceholder{}''\\
Passage: ``\promptPlaceholder{}''\\[0.15em]
Use one word to represent the passage in a retrieval task.
Make sure your word is in lowercase.\\
\promptImEnd{}\\[0.15em]
\promptImStart{}assistant\\
The word is ``\promptPlaceholder{}\\
\promptImEnd{}
\end{promptbox}

\begin{promptbox}{\ours{} target query-encoding prompt}
\promptImStart{}user\\
Query: ``\promptPlaceholder{}''\\[0.15em]
Given the few-shot examples above, imagine what a relevant passage
for this query would likely be about. Then use one lowercase word to
represent that expected passage in a retrieval task.\\
\promptImEnd{}\\[0.15em]
\promptImStart{}assistant\\
The word is ``
\end{promptbox}

\begin{promptbox}{\ours{} document-side in-context example}
\promptImStart{}system\\
You are an AI assistant that can understand human language.\\
\promptImEnd{}\\[0.15em]
\promptImStart{}user\\
Query: ``\promptPlaceholder{}''\\
Passage: ``\promptPlaceholder{}''\\[0.15em]
Use one word to represent the query in a retrieval task.
Make sure your word is in lowercase.\\
\promptImEnd{}\\[0.15em]
\promptImStart{}assistant\\
The word is ``\promptPlaceholder{}\\
\promptImEnd{}
\end{promptbox}

\begin{promptbox}{\ours{} target document-encoding prompt}
\promptImStart{}user\\
Passage: ``\promptPlaceholder{}''\\[0.15em]
Given the few-shot examples above, imagine what a relevant query for
this passage would likely be about. Then use one lowercase word to
represent that expected query in a retrieval task.\\
\promptImEnd{}\\[0.15em]
\promptImStart{}assistant\\
The word is ``
\end{promptbox}

\end{minipage}
\end{adjustbox}

\caption{Prompt templates used for PromptReps and \ours. The first
two panels show the query- and document-encoding prompts used by
PromptReps. The next two panels show the query-side in-context example
and target prompt used by \ours. The final two panels show the
corresponding document-side prompts. \ours (Zero-Shot) only utilizes the \ours target query-encoding and document-encoding prompts, without in-context examples.}
\label{fig:representation_prompts}
\end{figure*}

\end{document}